\documentclass[useAMS, usenatbib]{mnras}
\usepackage{amsmath,color}
\usepackage{slantsc}
\usepackage{graphicx}

\newcommand{\ms}{ h^{-1}{\rm M_{\odot}}}

\begin{document}
\bibliographystyle{mnras}
\graphicspath{{./figs/}}
\title[]{Projection effects in the strong lensing study of subhaloes }

\author[Ran Li et. al]
       {\parbox[t]{\textwidth}{
        Ran Li$^{1}$\thanks{E-mail:ranl@bao.ac.cn}, Carlos  S. Frenk$^{2}$, Shaun Cole$^{2}$, Qiao Wang$^{1}$, Liang Gao$^{1}$ 
       }
        \vspace*{3pt} \\
  $^{1}$Key laboratory for Computational Astrophysics, National Astronomical Observatories, Chinese Academy of Sciences, Beijing, 100012, China\\
  $^{2}$Institute for Computational Cosmology, Department of Physics, University of Durham, South Road, Durham, DH1 3LE\\
          }
\maketitle

\begin{abstract}
  The defining characteristic of the cold dark matter (CDM)
  hypothesis is the presence of a very large number of low-mass haloes,
  too small to have made a visible galaxy. Other hypotheses for the
  nature of the dark matter, such as warm dark matter (WDM), predict a
  much smaller number of such low-mass haloes. Strong lensing systems
  offer the possibility of detecting small-mass haloes through the
  distortions they induce in the lensed image. Here we show that the
  main contribution to the image distortions comes from haloes along
  the line of sight rather than subhaloes in the lens as has normally
  been assumed so far. These interlopers enhance the differences
  between the predictions of CDM and WDM models.  We derive the total
  perturber mass function, including both subhaloes and interlopers,
  and show that measurements of approximately 20 strong lens systems
  with a detection limit of $M_{\rm low} =10^7\ms$ would distinguish
  (at 3$\sigma$) between CDM and a WDM model consisting of 7~keV
  sterile neutrinos such as those required to explain the recently
  detected 3.5 ~keV X-ray emission line from the centres of galaxies
  and clusters.
\end{abstract}

\section{Introduction}

Under the cold dark matter (CDM) hypothesis, the power spectrum of
linear density perturbations has power on all scales down to a very
small cutoff which depends on the nature of the cold particles but is
typically of order the Earth mass \citep{Green2005}.  As a result, the
mass function of CDM haloes increases roughly as a power law to low
masses \citep{Springel2008,Diemand2008} and the defining characteristic
of a CDM universe is the existence of a very large number of low-mass
haloes. Most of these are too small for gas to have cooled in them to
form visible galaxies \citep[e.g][]{Efstathiou1992,Benson2002,Sawala2016}.

Alternative candidates for the dark matter such as sterile neutrinos
behave as warm dark matter (WDM). Their free streaming in the early
universe erases perturbations much larger than the Earth mass,
typically on the scale of dwarf galaxies. As a result, these models
predict far fewer small-mass haloes than CDM and none at all below the
corresponding cutoff mass in the power spectrum which also depends on
the properties of the particles.
\citep[e.g.][]{Avila-Reese2003,Lovell2012,Lovell2016,Kang2013,Schneider2012,Bose2016}. A
particularly topical candidate of this kind is a sterile neutrino of
mass of 7~kev whose decay could explain the 3.5 keV line recently
detected from the centres of galaxies and clusters
\citep{Bulbul2014a,Boyarsky2014a}. In the ``coldest'' example of a 7~kev
sterile neutrino, the cutoff occurs at a mass of a few times $10^8
$ $\ms$ \citep{Bose2016}.
 
CDM and viable WDM models predict similar numbers of faint dwarf
galaxies such as those observed as satellites around the Milky Way
\citep{Kennedy2014,Lovell2015}. Although the recent discovery of new
satellites may rule out some currently acceptable WDM models
\cite{Bose2016b}, a definitive test of WDM and indeed of CDM, requires
searching for the even smaller haloes which failed to make a galaxy
and thus remain dark. Analyses of strong lensing systems offers the
possibility of achieving exactly this: \cite{Koopmans2005} and
\cite{Vegetti2009a} showed that small haloes projected onto an
Einstein ring or giant arc cause a potentially detectable distortion
of the image and \cite{Vegetti2009b} showed that a Bayesian analysis
of sufficiently deep photometric data can be used to constrain the
subhalo mass function (SHMF) \citep[see also][]{Vegetti2012,
  Vegetti2014, Hezaveh2016}.


The technique proposed by \cite{Vegetti2009a} has already returned the 
detection of a halo of mass $1.9\pm 10^8$ $\ms$ in the Einstein ring of
JVAS B1938+666 \citep{Vegetti2012}. These authors claim that with imaging data
of similar quality the detection sensitivity can reach $2\times 10^7$
$\ms$. In a recent paper \citep{Li2016b},  we showed that
observations of approximately 100 strong lens systems with a detection
limit of $M_{\rm low }= 10^7 \ms$ could, in principle, distinguish CDM
for even the coldest 7~kev sterile neutrino dark matter model.  Of course,
failure to detect haloes of such low mass would conclusively rule out
CDM altogether. 

A common assumption made in studies of strong lensing is that the
haloes that perturb the image lie at the same redshift as the main
lens, i.e. that they are subhaloes of the lens.  However, it is possible that the
large number of haloes along the line of sight to a lens
could be the dominant source of distortion of the lensed image. Many
previous studies have shown that the line of sight haloes can play an important
role in the modeling of lensed quasar systems\citep[e.g.][]{Chen2003,Xu2009,Metcalf2005b,Wambsganss2005}.   In
this paper, we calculate the contribution of these ``interlopers'' and
investigate how they affect the prospects of distinguishing different
dark matter candidates.

Since the majority of the distortions are produced by dark haloes and
subhaloes, at first sight baryon effects may seem to be
irrelevant. However, this is not quite true: the visible galaxy at the
centre of the lens can, in principle, destroy dark subhaloes by
dynamical effects such as tidal stripping. In this paper 
we neglect baryon effects but we investigate those in a companion
paper using the APOSTLE hydrodynamic simulations \citep{Sawala2016c}. 

The paper is organized as follows. In Section~\ref{sec:num} we
calculate the number density of line of sight haloes both in CDM and
WDM models using the respective halo mass functions.  In
Section~\ref{sec:perturbing} we estimate the effect of individual
interlopers and derive the effective perturber mass functions.  In
Section~\ref{sec:constraint} we illustrate the constraining power of
halo/subhalo detection from multiple lens systems when including 
interlopers. Our conclusions are summarized in Section~\ref{sec:sum}.
 
\section{Number density of haloes along the line of sight}
\label{sec:num}

\begin{figure}
  \includegraphics[width=0.7\textwidth,angle=270]{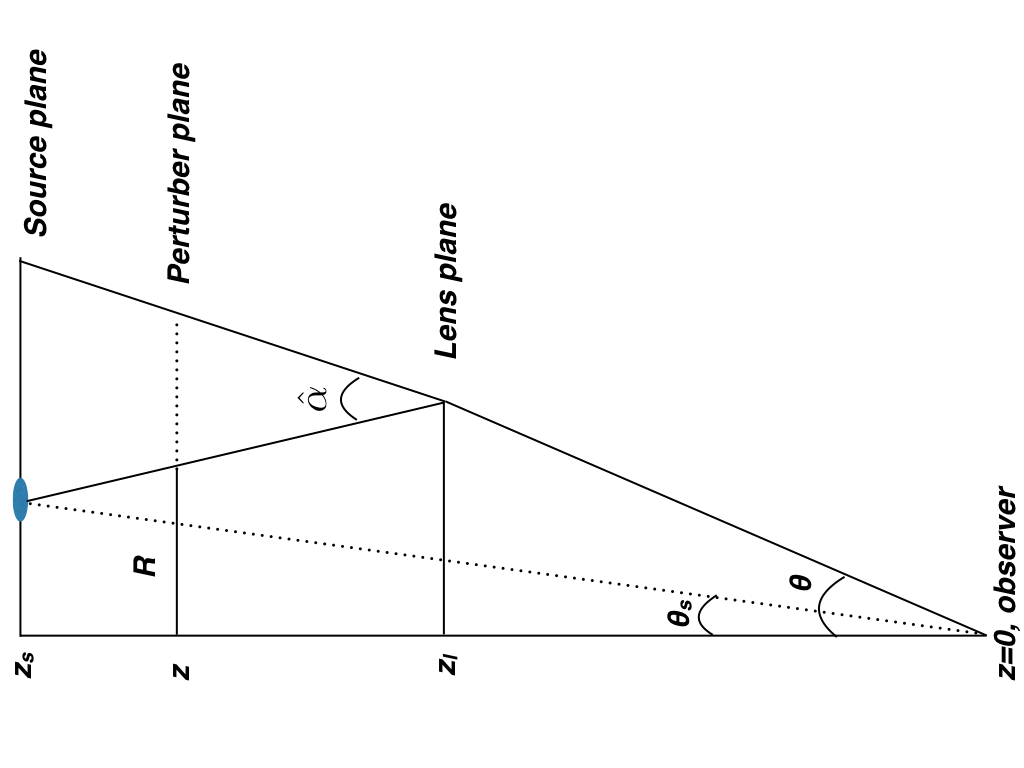}
  \caption{Geometry of the source/lens/observer system. $\theta_s$ is the
    position of the source and $\theta$ is the position of the
    image. $R(z,\theta)$ is the transverse distance corresponding to
    angle $\theta$ at redshift $z$.  When $z< z_l$, $R(z,\theta)$ is
    simply $D(0,z)\theta$, where $D(0,z)$ is the comoving
    distance from observer to redshift $z$; when $z>z_l$,
    $R(z,\theta)=D(0,z)\theta - \hat{\alpha}D(z_l,z)$ }
    \label{fig:sketch}
\end{figure}

Let $\theta_E$ be the Einstein radius of a lens. The light rays that
cross an annulus of thickness of $2\delta\theta$ around the lens form
a light cone whose volume is given by:
\begin{equation}
V=\int_0^{z_s} \pi \left[ R(z,\theta_E +\delta\theta)^2 - R(z,\theta_E -\delta\theta)^2\right]  \frac{d\chi(z)}{dz} dz \,,
\end{equation}
where $\chi(z)$ is the comoving distance from the observer at redshift 0
to redshift $z$, and $R(z,\theta)$ is the transverse distance corresponding to
angle $\theta$ at redshift $z$. When $z< z_l$, $R(z,\theta)$ is simply
$D(0,z)\theta$, where $D(0,z)$ is the comoving distance from
the observer to redshift $z$. When $z>z_l$, $R(z,\theta)=D(0,z)\theta
- \hat{\alpha}D(z_l,z)$ (see the sketch in Fig.~\ref{fig:sketch}), where
$\hat{\alpha}$ is the deflection angle of the lens.  Thus, 
$R(z,\theta)$ may be written as:
\begin{eqnarray}R(z,\theta)=
\begin{cases}
\theta D(0,z), &z<z_l\cr
 \theta D(0,z)-\hat{\alpha}D(z_l,z), & z>z_l
 \end{cases}
\end{eqnarray}
For a singular isothermal sphere (SIS) lens, $\hat{\alpha}= \theta_E
D(0,z_s)/D(z_l,z_s)$ is a constant.  The total number of
haloes in the light cone with mass in the range, $[M_1,M_2]$, is given
by:
\begin{equation}
N_{\rm los}(\theta_E,\delta\theta)=\int_0^{z_s} n(M_1,M_2,z) \frac{dV}{dz}dz \,,
\end{equation}
where
\begin{equation}
n(M_1<m<M_2,z)=\int_{M_1}^{M_2} \frac{dn(m,z)}{dm}dm \,,
\end{equation}
where $\frac{dn(m,z)}{dm}$ is the halo mass function at redshift $z$.

\begin{figure}
  \includegraphics[width=0.5\textwidth]{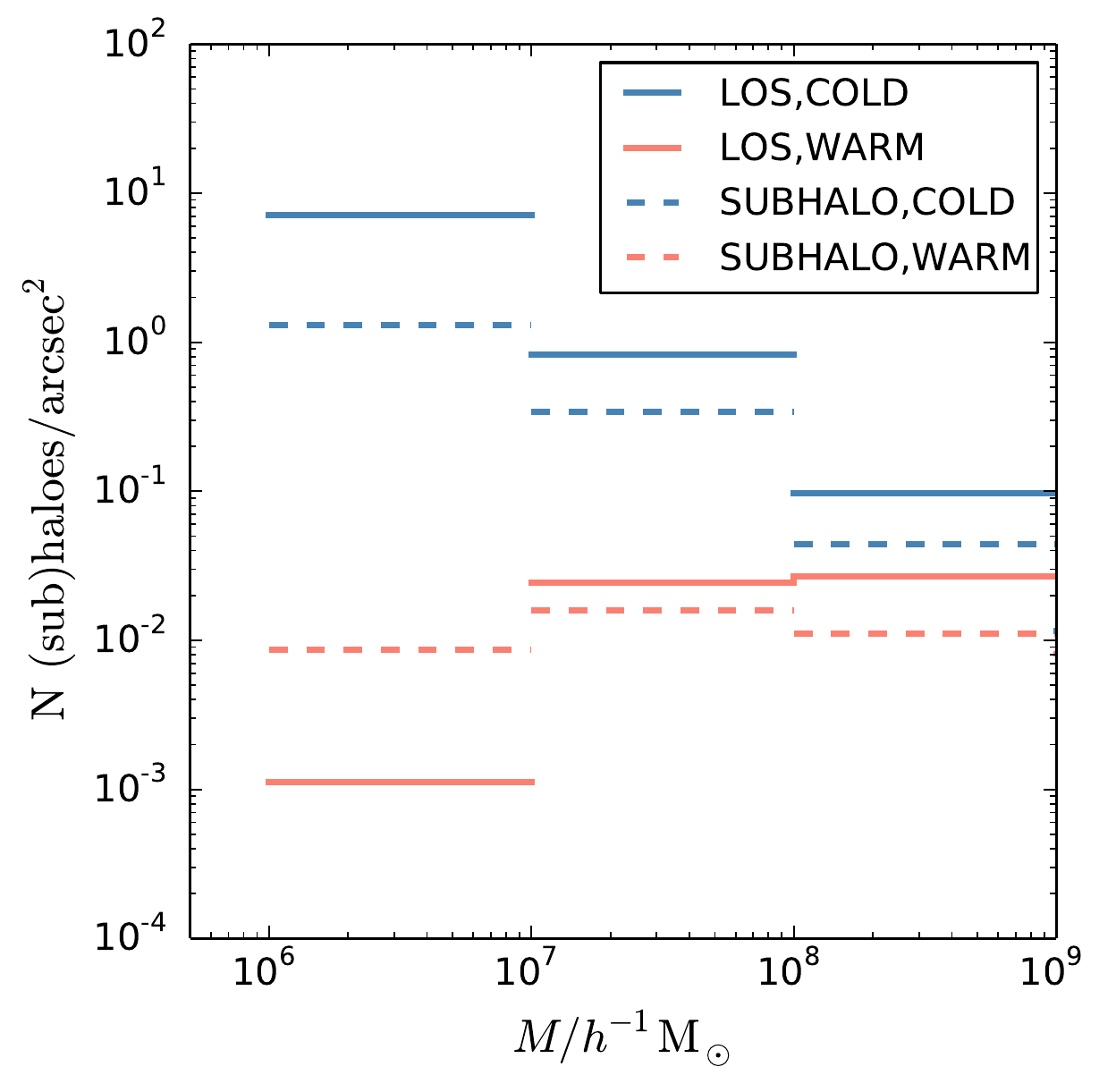}
  \caption{Relative contributions of lens subhaloes and interlopers. The
    dashed bars show the projected number density of subhaloes in the
    Einstein ring region of a lens in a halo of $10^{13}$~$\ms$
    at redshift $z_l=0.2$. The solid bars show the projected number density
    in that region of line of sight haloes. Blue and black lines are
    for CDM and red for WDM, as indicated in the legend. The width of the bars show the
    range of each mass bin. The projected subhalo number density is calculated with Eqn. 8 }
\label{fig:density}
\end{figure}

In Fig.~\ref{fig:density} we compare the projected number density of
interlopers and lens subhaloes in the Einstein ring region for a lens
in a CDM halo of mass $10^{13}$ $\ms$ at $z_l=0.2$.  To calculate
the number density of interlopers we use the formula for the halo mass
function proposed by \citep{sheth1999}. We use the projected
number density of subhaloes derived by \citet{Xu2015} from the Phoenix and Aquarius
N-body simulations \citep{Gao2011, springel2009}.  The projected
number density of line of sight dark matter haloes is larger than that
of subhaloes associated with the lens by a factor of 2-5.

The corresponding projected number densities for a WDM model are also
shown in Fig.~\ref{fig:density}. This model comes from the high
resolution \textsc{coco-warm} simulation, the WDM run of the
Copernicus Complexio project \citep{Hellwing2016,Bose2016,Bose2016b},
which corresponds to a thermal WDM particle of mass 3.3~keV. This is
indistinguishable from a sterile neutrino model of mass 7~keV with
leptogenesis parameter, $L_6=8.66$ which corresponds to the coldest
sterile neutrino model consistent with the dark matter decay
interpretation of the 3.5~keV X-ray line \citep{Lovell2016}. Ruling
out this extreme model would exclude the entire family of 7~keV
sterile neutrinos.


\section{The number density of perturbing interlopers}
\label{sec:perturbing}

The lensing effect of an interloper halo projected onto the Einstein
ring of a lens depends on its mass, structure and redshift. There is a
strong degeneracy between mass and redshift. In this section we
calculate the lensing effect of interlopers by creating mock Einstein
ring images using ray-tracing simulations. The mass model for the lens plus perturber
consists of a main lens and a halo along the line of sight. 

We assume that the main lens is at $z_l=0.2$ and has a SIS profile with
 $\sigma_v=350$ $\rm km~s^{-1}$. An interloper of mass $M_{\rm int}=5\times
10^6\ms$ is placed at $z_{\rm int}=0.18$ and a Gaussian source is place at redshift $z_l=1.0$. 
The perturber density profile is assumed
to have the NFW form \citep{NFW97} with concentration given by the
median of to mass-concentration relation of \citet{Neto2007}. The brightness
distribution of the source galaxy is assumed to be Gaussian with
dispersion $\sigma_{\rm source}=0.05\arcsec$.

We then use a ray-tracing code to generate a lensed image
  on a plane of 500$\times$500 pixels. The size of each pixel is
  0.043", which is close to the Hubble telescope imaging resolution.
  We assume that the uncertainty in the flux in each pixel is 10\% of
  the mean flux. Once a mock image has been generated, we use an MCMC
  minimization method to fit the image with the same mass model as
  above in which all parameters, except the redshift, $z_{\rm int}$,
  and mass, $M_{\rm int}$, of the interlopers are fixed.  

\begin{figure}
  \includegraphics[width=0.5\textwidth]{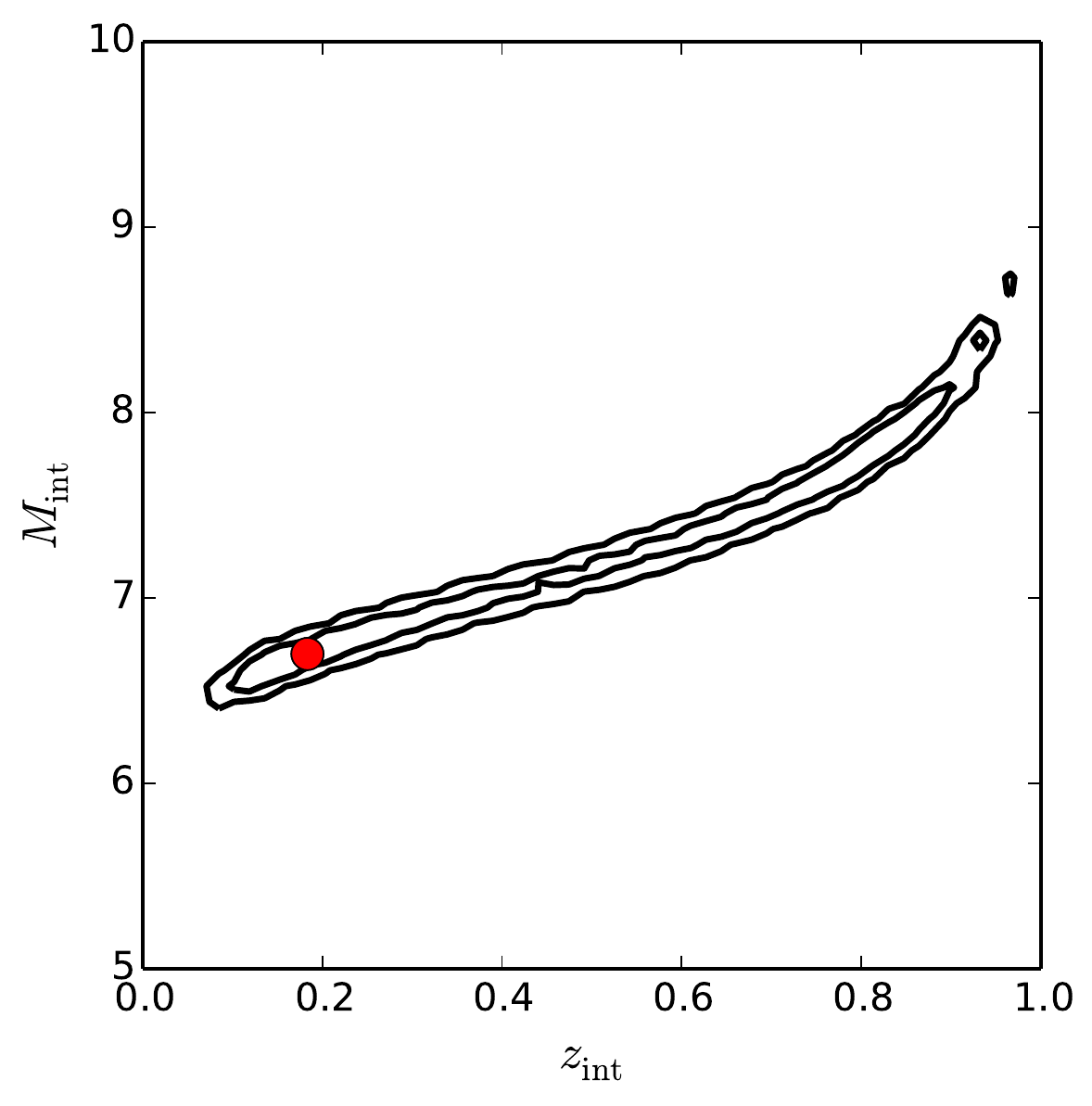}
  \caption{The posterior distribution of interloper mass, $M_{\rm
      int}$, and redshift, $z_{\rm int}$ for an interloper of mass of
    $5\times10^6\ms$ placed at $z=0.18$.  The contours show the 68\%
    and 95\% confidence levels. There is a clear
      degeneracy between $z_{\rm int}$ and $M_{\rm int}$; we have
      assumed a flux uncertainties of 10\% of the mean flux. }
 \label{fig:degeneracy}
\end{figure}

The posterior distribution of $z_{\rm int}$ and $M_{\rm int}$,
displayed in Fig.~\ref{fig:degeneracy},  clearly shows that these two
parameters are highly degenerate with the 1-$\sigma$ contour including
a very wide range of redshifts and masses. A low mass interloper in
front of the lens can have a similar lensing effect as a higher mass
interloper behind the lens. Since the halo mass function increases
with decreasing mass, the higher the lens redshift, the more important
interlopers become.  We find that within a small range of the Einstein radius, the
form of the degeneracy between $z_{\rm int}$ and $M_{\rm int}$ is independent
of the angular position of the interloper.

If we fix the interloper redshift to be $z_{\rm lens}$ during the
fitting, we can derive a best-fit ``effective mass,'' $M_{\rm
  eff}(M_{\rm int},z_{\rm int},\theta)$, where $M_{\rm int}$ and
$z_{\rm int}$ are the true mass and redshift of the interloper, and
$\theta$ is its angular position. In other words, we can use a subhalo
of $M_{\rm eff}$ at redshift $z_{\rm lens}$ to model an image
perturbation caused by an interloper of mass $M_{\rm int}$ at redshift
$z_{\rm int}$.

Note that, in this ray-tracing test, we do not include
  observational details such as the PSF, source complexity or noise
  variation across the the image.  If an interloper with mass, $M_{\rm
    int}$, and redshift, $z_{\rm int}$, has the same lensing effect as
  a subhalo of mass, $M_{\rm eff}$, in this idealized simulation, it
  should have the same lensing effect as a subhalo of $M_{\rm eff}$ in
  a more realistic simulation with PSF and noise added. Our idealized
  ray-tracing simulation therefore encapsulates the key information
  regarding the mass-redshift degeneracy inherent in the modelling of
  the perturber. 

We assume that the mass detection limit for subhaloes is $M_{\rm low}$
within a thin annulus of thickness $\delta\theta$ around the Einstein
radius.  Then, any interlopers with $M_{\rm eff}> M_{\rm low}$ can
be detected. We refer to these as ``perturbing interlopers".  The
projected number density of perturbing interlopers can be written as:
\begin{equation}
\begin{aligned} 
&\Sigma_{\rm pb}(>M_{\rm low}) = \\
 &\frac{1}{\theta_E\delta\theta} 
 \int_0^{z_s} n(M_{\rm int, low}<m<M_{\rm max},z) R(z,\theta_E)\delta R(z,\delta \theta)\frac{d\chi(z)}{dz}dz \,,
\end{aligned}
\end{equation}
where $M_{\rm int, low}$ is defined implicitly by $M_{\rm eff}( M_{\rm int, low}, z)=M_{\rm low}$, $M_{\rm
  max}=10^{11}\ms$ is a cut off mass we impose for the maximum halo mass considered for the mass
function in the volume along the line of sight to the lens.  The
number density of haloes is dominated by the low mass end, the exact choice
of $M_{\rm max}$ has no effect on the results.

The projected subhalo number density in a CDM universe can be written as: 
\begin{equation}
  \Sigma_{\rm sub,cdm}(m)=\Sigma_0 \,\left(\frac{m}{\ms} \right)^{-\alpha} \,,
\label{eq:dn_cdm}
\end{equation}
and the cumulative surface density of subhaloes in the mass range,
$[M_1,M_2]$,  can be written as
\begin{displaymath}
 \Sigma_{\rm sub,cdm}(M_1<m<M_2)= \left \{ \begin{array}{ll}
        \Sigma_0 \ln{\frac{M_2}{M_1}} & \textrm{if  $\alpha= 1$}    \,, \\
 \\
 \frac{\Sigma_0}{1-\alpha}(M_2^{1-\alpha} - M_1^{1-\alpha}) & \textrm{otherwise} \\
    \end{array} \right . \,. 
  \label{eq:n_cdm}
\end{displaymath}

According to Fig.~\ref{fig:density}, the projected number density of
subhaloes with mass $10^6$ to $10^7$ $\ms$ is 1.3/$\rm
arcsec^2$. Thus, we have $\Sigma_0=3.1 \times 10^5 / \rm arcsec^2$, assuming $\alpha=1.9$.

Following \cite{Schneider2012} and \cite{Lovell2014}, we write the subhalo mass function as 
\begin{equation}
\frac{d\Sigma_{\rm sub, wdm}}{dm}=\frac{d\Sigma_{\rm sub, cdm}}{dm}(1+m_{\rm c}/m)^{-\beta} \,.
\label{eq:dn_wdm}
\end{equation}
and the cumulative mass function, $\Sigma_{\rm sub, wdm}(M1,M2)$, can be written as:
\begin{equation}
\Sigma_{\rm sub, wdm}=  \frac{\Sigma_0}{1-\alpha+\beta}\left[F(M_2,\alpha,\beta,m_c)-F(M_1, \alpha,\beta,m_c) \right] \,,
\label{eq:n_wdm}
\end{equation}
where,
\begin{equation}
F(x,\alpha,\beta,m_c)=\frac{x^{1-\alpha+\beta} }{m_c^{\beta}} {\rm
 _2F_1} \left( \beta,1-\alpha+\beta,2-\alpha+\beta,\frac{-x}{m_c} \right) \,. 
\end{equation}
Here $\rm _2F_1$ is the hypergeometric function. In the \textsc{coco-warm} simulation, $\beta=1.3$ and $m_{\rm c}=1.3\times10^8$ $\ms$ \citep{Li2016b}.  

\begin{figure}
  \includegraphics[width=0.5\textwidth]{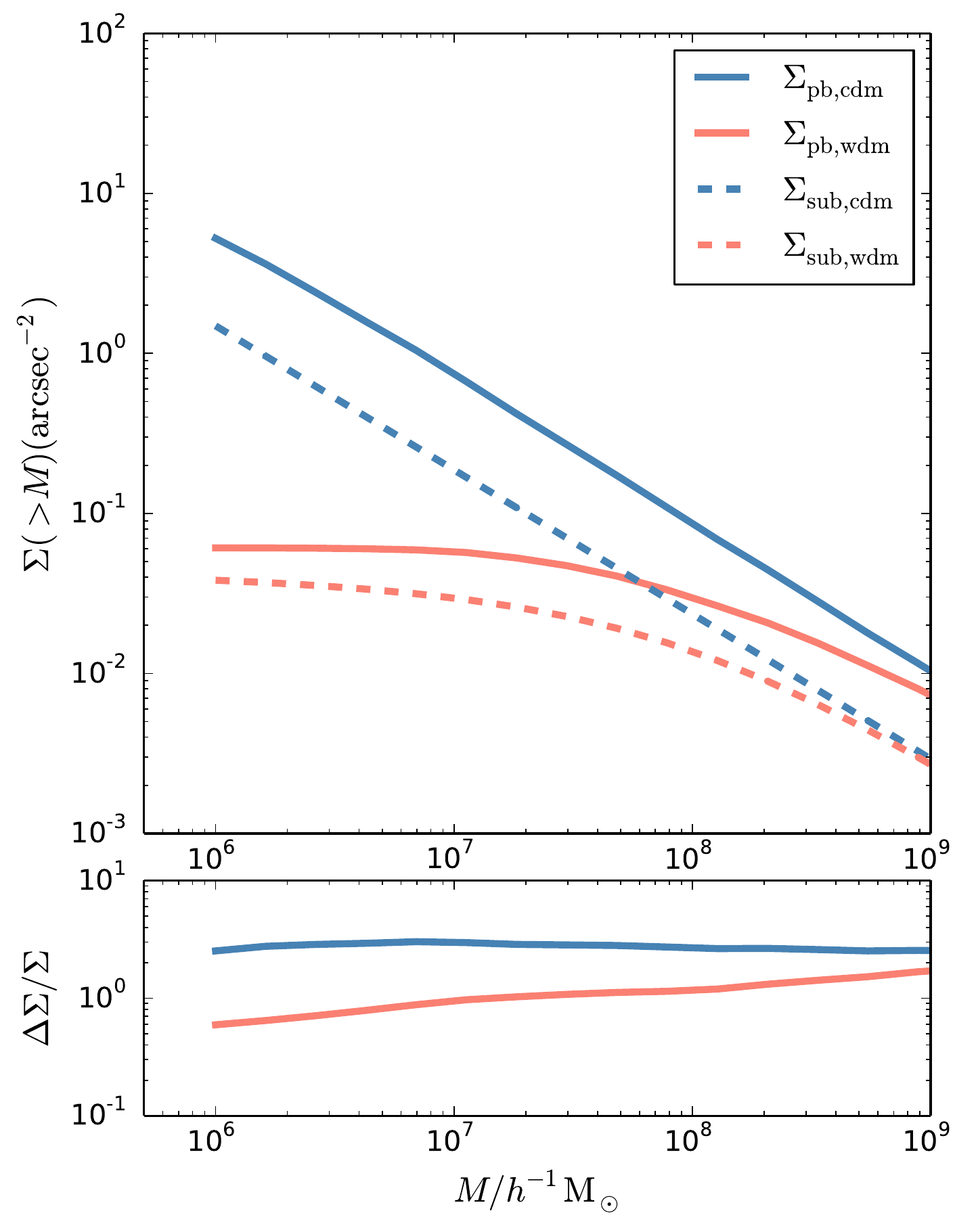}
  \caption{Cumulative number densities of perturbing interlopers and subhaloes
    and as a function of the subhalo mass detection limit.  The blue
    dashed line gives the number density of subhaloes of mass greater than
    $M_{\rm low}$ in a host halo of mass of $10^{13}\ms$ at redshift
    0.2, derived by \citealt{Xu2015}. The red dashed line gives the
    corresponding number density of subhaloes in the
    \textsc{coco-warm} simulation.  The blue and red solid lines show
    the number density of perturbing interlopers along the
    line of sight in the CDM and \textsc{coco-warm} cases
    respectively. The lower panel shows $[\Sigma_{\rm pb}(>M)-\Sigma_{\rm sub}(>M)]/\Sigma_{\rm sub}(>M)$.}
\label{fig:perturbing}
\end{figure}

Fig.~\ref{fig:perturbing} shows the cumulative number
  density of perturbing interlopers in CDM and in the WDM model
of \textsc{coco-warm}.  For comparison, we overplot the
projected number density of subhaloes in a host halo of $10^{13}\ms$
at redshift $z_l=0.2$. Fig.~\ref{fig:perturbing} shows that for both
CDM and WDM the perturbing interlopers dominate the distortions in the
Einstein ring image.  In the CDM case, the projected
  number density of perturbing interlopers is $\sim 3$ times the
  number density of lensing subhaloes.  In the \textsc{coco-warm}
  case, the excess is a factor of 2 at $M=10^9$ $\ms$, but decreases
  to ~50\% at $M=10^6$ $\ms$. Thus, the interlopers act to magnify the
  difference in the number of detectable perturbers in the two cases.
  The diffferent boost factors between CDM and WDM are due to the
  differences in the shapes of the halo mass function in the two
  models. The total perturber mass function is an integral over all
  perturbing interlopers and perturbing subhaloes.  In a CDM universe,
  the mass function over the mass range of interest follows a power
  law whose index is very similar to that of the subhalo mass
  function; the total perturber mass function is then just boosted by
  a constant factor.  The halo and subhalo mass functions in the WDM
  model both have a mass cutoff at about $10^8$ $\ms$. The detection
  limit for interloper haloes varies with redshift so, in this case,
  the total perturber mass function does not have exactly the same
  shape as the subhalo mass function.  

\section{Constraints on the identity of dark matter}
\label{sec:constraint}

As we have seen, to predict correctly the distortions in the Einstein
ring image it is necessary to model the total perturber surface
density, $\Sigma_{\rm tot}=\Sigma_{\rm sub}+ \Sigma_{\rm pb}$, which
includes both interlopers and subhaloes in the lens.  We find that the
total surface density of perturbers in the CDM and WDM models can be
described by a formula of the form of Eqn.6 and
Eqn.7 respectively.  For CDM, $\Sigma_{\rm tot}$ can be used with $\alpha=1.9$ and $\log{\Sigma_0}=6.2$; for
\textsc{coco-warm} , $\Sigma_{\rm tot}$ can be used with $\alpha=1.9$,
$\log{\Sigma_0}=6.1$, $\beta=1.3$ and $\log{(m_c/\ms)}=8.3$.  
 We can then exploit the difference
 in the perturber mass functions to attempt to constrain the identity of dark matter. 
 Here the key parameter is $m_c$ which describes the cut off mass for  the perturber mass function in the WDM model.

To explore the constraining power of a detection of strong lensing
perturbations we adopt a similar methodology to that introduced by
\citet{Li2016b}. First, we generate mock subhalo detections using the following
Monte Carlo method. 

We fix the lens and source redshifts to be $z_l=0.2$ and
  $z_s=1.0$ respectively, and assume that the lens galaxy is a SIS
  with velocity dispersion, $\sigma_v=350$ $\rm km~s^{-1}$, which is
  similar to those of the most massive lenses in the Sloan Lens ACS
  Survey\citep[SLACS,][]{Bolton2006} lens sample. 

For each lens, we randomly sample subhaloes and perturbing interlopers around the
Einstein ring region according to their mass functions.

Following \citet{Li2016b} and \citet{Vegetti2009b}, we assume that
only perturbers that fall in a thin annulus around the Einstein radius
of width $2\delta \theta=0.6$ arcsec can be detected. We consider two
different detection limits, $M_{\rm low}$: $10^8$ $\ms$, the best
current limit using HST imaging \citep{Vegetti2014}, and $10^7$ $\ms$,
the detection limit that can be reached using  Laser Guide
  Star (LGS) Adaptive Optics (AO) imaging with Keck
\citep{Vegetti2012} or a next generation telescope like the TMT, or
VLBI \citep{Skidmore2015,McKean2015}. We assume that each subhalo
detection has a Gaussian measurement error with standard deviation,
$\sigma_m=M_{\rm low}/3$.  We generate two sets of mock
  detection catalogues,  with the mass functions appropriate to CDM and
  \textsc{coco-warm} respectively.

 We then perform an MCMC fit for each mock detection
  catalogue. There are four free parameters in the model: $\alpha$,
  $\Sigma_0$, $\beta$ and $m_c$. Given these model parameters, the
  mean number of detected subhaloes can be written as:
\begin{eqnarray}
\begin{split}
&\mu(\alpha,\beta,m_{\rm c},\Sigma_0)= 4\pi\theta_{\rm E}\delta\theta \\
& \int^{\infty}_{M_{\rm low}} \int^{M_{\rm max}}_{M_{\rm min}} \frac{d\Sigma_{\rm tot}}{dm}\frac{1}{\sqrt{2\pi}\sigma_m}\exp \left[\frac{-(m-m')^2} {2\sigma_m^2 }\right] dm' dm \,. \\
 \end{split}
\end{eqnarray}

The likelihood of finding a set of $n_s$ subhaloes of masses, ${\bf m}\equiv
\{m_1,m_2,\dots m_{n_s}\}$, in one Einstein ring system is then given by:
\begin{equation}
\mathcal{L}(n_s,{\bf m} | {\bf p, q})=\frac{e^{-\mu} \mu^{n_s}}{n_s!}\prod^{n_s}_{i=1}P(m_i|{\bf p, q})\,,
\end{equation}
where the vector, ${\bf p}=\{\Sigma_0,\alpha, \beta,m_c\}$, contains
the parameters of the model and the vector, ${\bf q} =\{ M_{\rm min},
M_{\rm max}, M_{\rm low} \} $, contains the values of the parameters
that define the minimum and maximum masses cut off we consider for the perturber mass function
and the mass detection limit. The parameters, ${\bf
  q}$, are fixed during the fitting process. In this process, we set
$M_{\rm min}=10^6$, and $M_{\rm max}=10^{11}$ $\ms$. The exact
choice of $M_{\rm min}$ and $M_{\rm max}$ does not affect the results.

The term $P(m_i|{\bf p, q})$ gives the probability density of detecting a
subhalo of measured mass, $m_i$:
\begin{equation}
P(m_i|{\bf p, q})=\frac{\int^{M_{\rm max}}_{M_{\rm min}} \frac{d\Sigma_{\rm tot}}{dm}\exp\left[\frac{-(m_i-m')^2}{2\sigma_m^2 }\right] dm' } 
{\int^{M_{\rm max}}_{M_{\rm low}} \int^{M_{\rm max}}_{M_{\rm min}} \frac{d\Sigma_{\rm tot}}{dm} \exp \left[\frac{-(m-m')^2}{2\sigma_m^2 }\right] dm' dm}
\end{equation}
The denominator in this equation is a normalization factor.  The total
likelihood for $N$ lenses may be written as:
\begin{equation}
\mathcal{L}_{\rm tot}=\prod_{j=0}^{N}\mathcal{L}(n_{j},{\bf m_j}|\bf p,q) \,,
\end{equation}
where $n_{j}$ and ${\bf m}_j$ are the number and masses of the perturbers
detected in the $j$th system.

Following \citet{Li2016b}, we adopt a Gaussian prior for $\alpha$ with
expectation 1.9 and standard deviation 0.1, and a Gaussian prior for
$\beta$ with expectation 1.3 and standard deviation 0.1. 
  We adopt flat priors for $\log(\Sigma_0)$ and $\log(m_c/\ms)$ in the
  ranges  $[1,10]$ and $[4,11]$ respectively. 

\begin{figure}
  \includegraphics[width=0.5\textwidth]{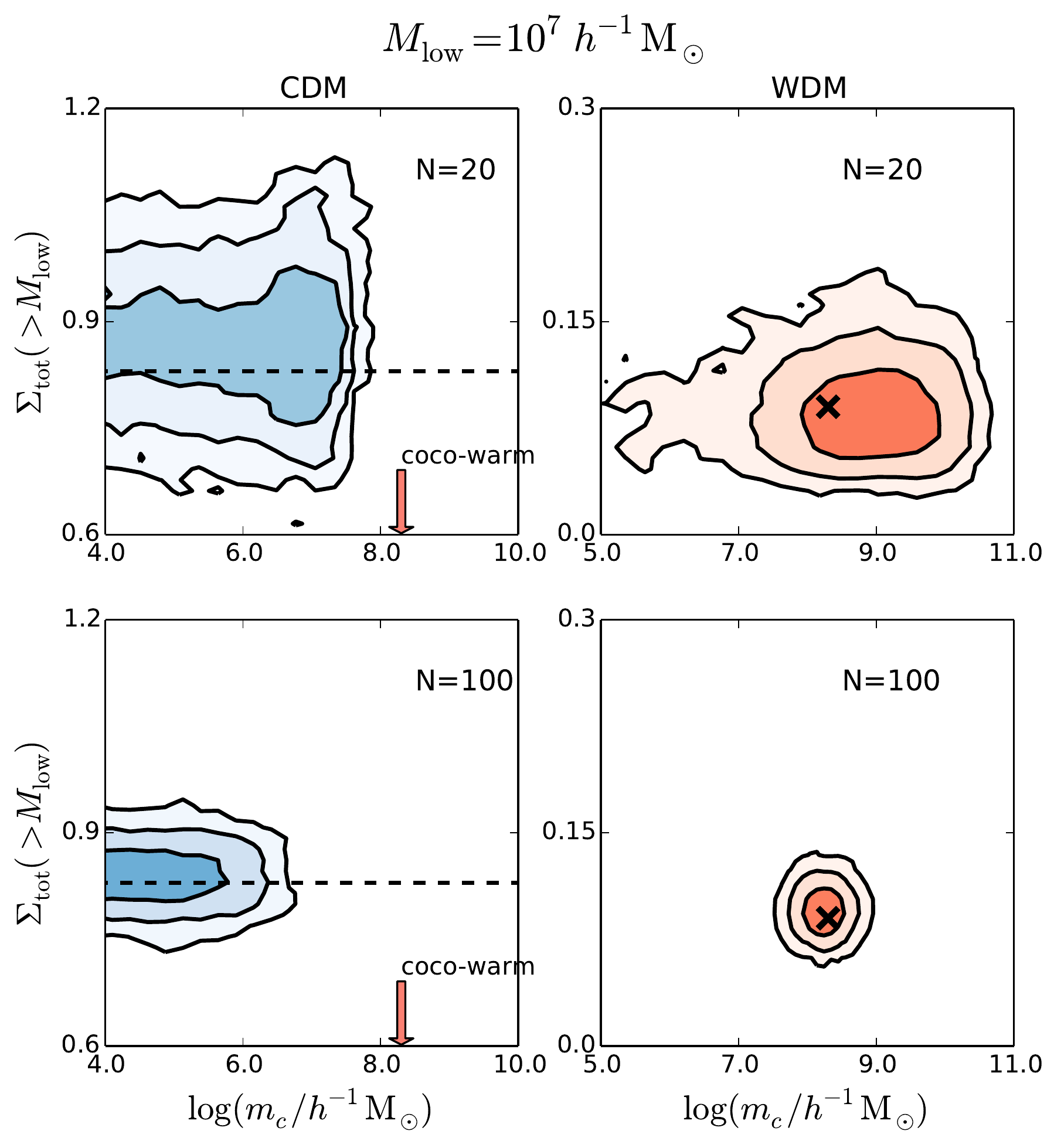}
  \caption{The posterior distribution of $\Sigma_{\rm tot}(>M_{\rm
      low})$ (in units of $\Sigma_{\rm tot}$
      arcsec$^{-2}$) and perturber mass function cutoff mass, $m_{\rm
      c}$. The contours indicate the 67\%, 95\% and 99.7\% confidence
    levels. The left panels show results for CDM while the right
    panels show results for the \textsc{coco-warm} model. The upper
    panels are for $N=20$ and the lower panels for $N=100$ lenses. The
    detection limit is assumed to be $M_{\rm low}=10^7\ms$. 
       In the right panels the crosses show the input
      values of $m_c$ and $\Sigma_{\rm tot}$; in the left panels the
      dashed lines indicate the input value of $\Sigma_{\rm tot}$.
      The arrows mark the value of $m_c$ for the \textsc{coco-warm}
      model.   }
 
 \label{fig:mcmc1}
\end{figure}

Fig.~\ref{fig:mcmc1} shows the posterior distributions of
  $\Sigma_{\rm tot}(>M_{\rm low})$, and the cutoff mass, $m_{\rm
    c}$. The upper panels are for $N=20$ lenses and the lower panels
  for $N=100$ lenses. In both cases the detection limit is assumed to
  be $M_{\rm low}=10^7\ms$.  The left panel shows the result for our
  CDM mock catalogues and the right panel for the \textsc{coco-warm}
  case.

Encouragingly, we find that a detection limit of $10^7$ $\ms$ is
sufficient to distinguish between the two dark matter models.
 If we live in a CDM universe (left panel), with a sample
  of only 20 lenses we are able to rule out $\log(m_c/\ms)=8.3$ at the
  $3\sigma$ level. By contrast, if we live in a universe in which the
  dark matter consists of 7 kev sterile neutrinos (right panel), with
  N=20 lenses and $M_{\rm low}=10^7 h^{-1}$ $\ms$ we can rule out, at
  the $3 \sigma$ level, all dark matter models with $\log {(m_c/\ms)}
  < 5$, which, of course, includes CDM! The constraining power
  increases with the number of lens systems. If the number is 100, and
  the dark matter is as in \textsc{coco-warm}, we can rule out all
dark matter models with $\log{(m_c/\ms)}< 7.5$ at $3\sigma$.

\begin{figure}
  \includegraphics[width=0.5\textwidth]{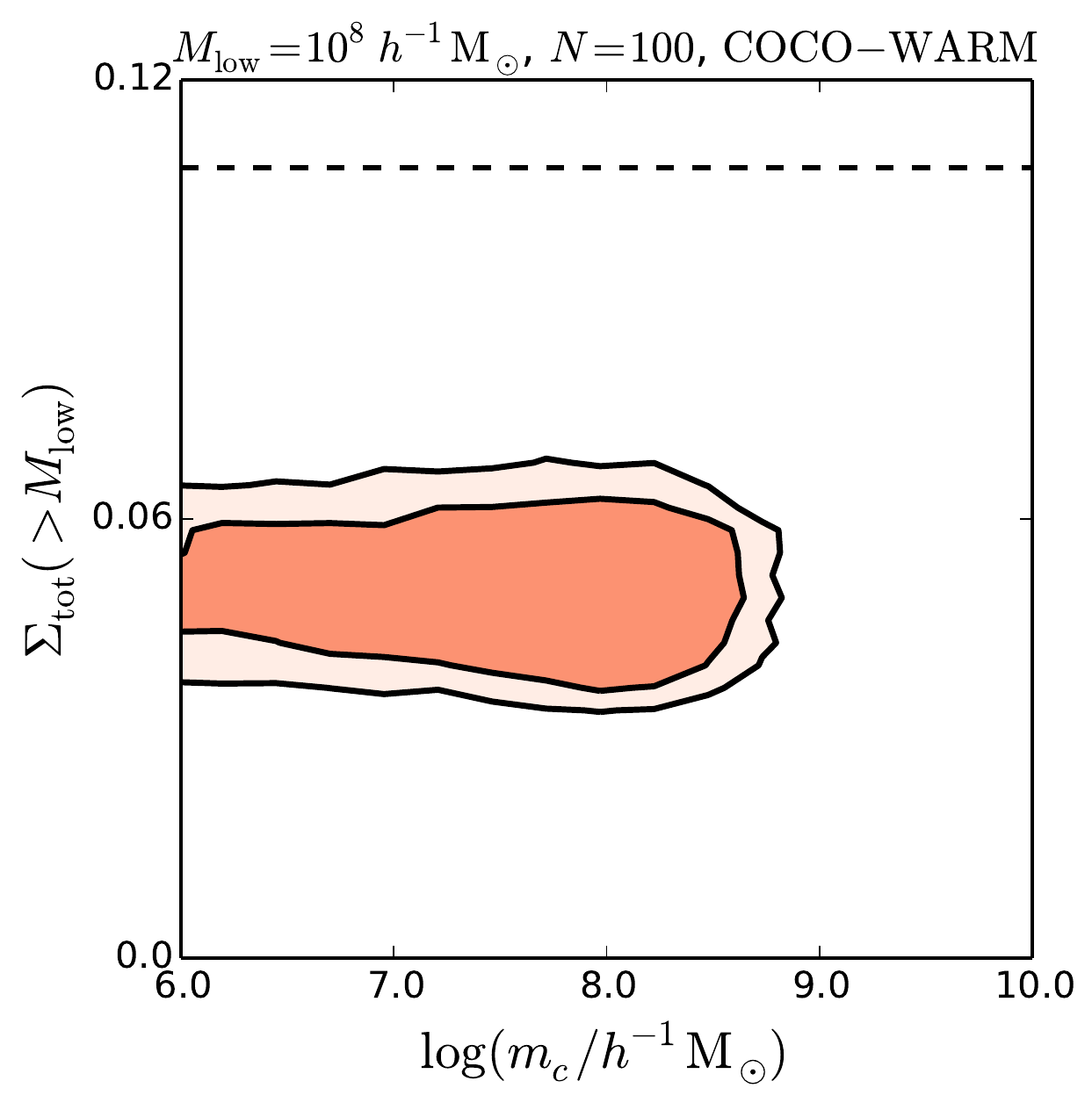}
  \caption{ As Fig.~\ref{fig:mcmc1}, but for a detection threshold,
    $M_{\rm low} =10^8 \ms$, and a number of lenses, $N=100$.  The
    input perturber mass function is from the \textsc{coco-warm}
    model. The black dashed line shows the expected value of
    $\Sigma_{\rm tot}$ in CDM. }
 \label{fig:mcmc2}
\end{figure}

Fig.~\ref{fig:mcmc2} shows the constraints on $m_c$ and $\Sigma_{\rm
  tot}$ that can be obtained for the \textsc{coco-warm} model with
$N=100$ and $M_{\rm low} = 10^8$ $\ms$. Dark matter models with $m_c >
10^9 \ms$ are disfavoured, but the CDM model cannot be ruled out by
this experiment. This agrees with the conclusion of Li16 that the
constraining power on $m_c$ is weaker when the detection limit,
$M_{\rm low} >10^8\ms$. Above this mass, the slope of the mass
function of perturbers in the \textsc{coco-warm} model is
intrinsically similar to that of CDM. On the other hand, with $N=100$,
one can place a tight constraint on $\Sigma_{\rm tot}$ which would provide
a strong hint that the dark matter is not CDM since, as we can see in
the figure, the best-fit $\Sigma_{\rm tot}$ is far below the prediction
of a CDM universe. This demonstrates that the identity of the dark
matter can be strongly constrained by the the total number of perturbations
alone.

 In this paper we have assumed a lens model with
  $\sigma_v=350$ km/s and $z_s=1$, which is near the upper envelope of
  the SLACS sample. More massive lenses, in combination with more
  distant sources, produce larger Einstein rings and these lead to larger
  volumes for interloper detection. These lenses should be high
  priority targets for future high-resolution observations. If we
  adopt a configuration similar to the average SLACS sample, with
  $z_l=0.7$ and $\sigma_v=275$ km s$^{-1}$ \citep{Bolton2006}, we
  require ~50\% more lenses to achieve similar constraining power.

  An important simplification we have made is to assume
  a uniform detection limit for the perturber over the entire Einstein
  ring. In a real situation the detection limit varies across
  the Einstein ring region and a sensitivity map that specifies the
  subhalo mass detection limit at each pixel of the image, like those
  made by \citet{Vegetti2014}, is crucial for constraining the
  perturber mass function.  Once such a map has been constructed, the
  strategy used in this paper can be applied with minor changes. In
  particular, in eqns.~11-13, one should first calculate the
  likelihood of detecting $n_s$ perturbers in the $i$th pixel of the
  $j$th lens and then sum the likelihood over all the pixels of all
  lenses.

  In this study we have also neglected the effects of the galaxy in
  the lens on the population of subhaloes orbiting in the same
  halo. In a recent paper, \citet{Sawala2016c} calculated the changes
  in the abundance and spatial distribution of subhaloes in the mass
  range $10^{6.5}$ to $10^{8.5}$ $\ms$, in haloes of mass of $10^{12}
  \ms$, caused by interaction with the central galaxy. By comparing
  the hydrodynamical simulations of the Apostle project of Local Group
  simulations with their dark matter only counterparts, they found the
  reduction in the number of subhaloes as a function of radial
  distance due to tidal disruption in the potential well deepened by
  the presence of the central galaxy to be approximately independent
  of subhalo mass.  At halocentric distances $r<50$~kpc, the
  number of subhalos is reduced by $\sim 40-50$\% and at radii in the
  range $r = 50-200$~kpc by 23\%.

  The host haloes in the Apostle simulations are an order of magnitude
  less massive than the halos we are considering in this study. If we
  assume that the reduction in numbers scales with $r/r_{\rm 200}$, we
  should expect the number of subhaloes in strong lenses system also
  to be $\sim20-50$\% smaller than the number predicted in dark matter
  only simulations. This effect, however, does not alter the
  conclusions in this paper because, as we have seen, the perturbers
  of Einstein ring systems are predominantly field haloes along the
  line of sight to the lens, rather than subhaloes. 

\section{Summary}
\label{sec:sum}

The most direct, and potentially conclusive, test of different models
for the dark matter is to measure the mass function of dark matter
halos in the low-mass regime where different models that agree with
CMB and large-scale structure data can be expected to differ.
Unfortunately, attempts to infer the small-mass end of the dark halo mass 
function from observations of visible galaxies are hampered by the
intrinsically low luminosity of faint objects and further complicated
by uncertainties in modeling baryon effects.

In contrast Einstein rings (and giant arcs) produced by strong
gravitational lensing, offer a clean and powerful means to 
detect small halos and measure or constrain the halo mass function.
These small halos perturb lensed images and by modelling these
perturbations, it is possible to detect individual haloes projected
onto the image and measure their mass. There is a strong degeneracy
between the mass of a perturber and its redshift. As a result, the
lensing effect of an interloper halo along the line of sight can be
modelled as that produced by a (sub)halo of some effective mass
located at the redshift of a lens.

In this paper we have compared the CDM model with a WDM model whose
linear perturbation power spectrum is that of a thermally produced
3.3~keV particle and provides a very good approximation to the linear
power spectrum of the coldest possible 7~keV sterile neutrino
consistent with a particle decay interpretation of the recently
discovered 3.5~keV line in the X-ray spectra of galaxies and
clusters. Ruling out this model by detecting small halos below the
cutoff mass in its predicted halo mass function would rule out all
7~keV sterile neutrino models. Similarly, a failure to detect
small-mass subhalos would rule out CDM. 

For both CDM and WDM models we have calculated the projected number
density of interlopers and compared it to the projected number of
subhaloes. We defined the `perturbing' interlopers as those that generate a
larger lensing signal than a subhalo of mass, $M_{\rm low}$.  We then
derived the effective mass function of perturbers, including both
perturbing interlopers and subhaloes. We find that the total number density of 
perturbers is 4 times of that of subhaloes in CDM and 1.5-2 times of that of subhaloes
 in our WDM model. Interlopers therefore boost the probability of
detection and act to magnify the difference between CDM and WDM.

 We find that a measurement of
only 20 strong lensing systems with a detection threshold of $M_{\rm
  low}=10^7$ $\ms$ is enough to distinguish between CDM and our WDM
model at the 2$\sigma$ level. With a survey of 100 strong lenses the
confidence level increases to 3$\sigma$. If the threshold mass,
$M_{\rm low}=10^8$ $\ms$, the constraint on the cutoff halo mass of
our warm dark matter model, $m_c$ becomes weaker because the slope of
the effective mass function above $10^8$ $\ms$ in this model is
similar to that in CDM but the constraint on the total number density
of perturbers is tight, thus retaining discriminating power between
the models.

Strong gravitational lensing provides, in principle, a clean test of
dark matter models. The quality of existing data and analysis
technique is already sufficient to detect dark low-mass haloes, too
small to have made a galaxy. As a result this technique is almost
unaffected by uncertain baryon effects, except for the possibility
that the disruption of subhaloes orbiting within a large halo may be
enhanced by the concentration of mass induced by the central
galaxy. This kind of processes can be quantified with hydrodynamic
simulations and, once this is achieved, as we shown a conclusive test
of the nature of the dark matter will be possible.  In particular
lensing measurements forthcoming in the next few years offer the
possibility of ruling out the main current candidates for the dark
matter, CDM and WDM.

 \section*{Acknowledgements}

 RL acknowledges NSFC grant (Nos.11303033,11511130054，11333001), support from
 the Newton Fund, Youth Innovation Promotion Association of
 CAS and YIPA of NAOC. CSF and SMC acknowledge the European Research Council Advanced
 Investigator grant, GA 267291, COSMIWAY. LG acknowledges support from the NSFC grant (Nos 11133003, 11425312), the Strategic Priority Research Program The Emergence of Cosmological Structure of the Chinese Academy of Sciences (No. XDB09000000), and a Newton Advanced Fellowship, as well as the hospitality of the Institute for Computational Cosmology at Durham University. 
 This work was supported by the Consolidated Grant [ST/L00075X/1] to Durham from the Science and
 Technology Facilities Council. This work used the DiRAC Data Centric
 system at Durham University, operated by the Institute for
 Computational Cosmology on behalf of the STFC DiRAC HPC Facility
 (www.dirac.ac.uk). The DiRAC system is funded by BIS National
 E-infrastructure capital grant ST/K00042X/1, STFC capital grant
 ST/H008519/1, STFC DiRAC Operations grant ST/K003267/1, and Durham
 University. DiRAC is part of the National E-Infrastructure.
 We thank the participants of the workshop ``Dark matter on the smallest scale" (Lorentz Centre -- 4-8 April 2016) for lively discussions and Simon White for his insightful comments.

\bibliography{ref}

\end{document}